\documentclass[preprint]{aastex}

\newcommand{\com}[1]{}

\usepackage{amsmath,amssymb,amsfonts}
\usepackage{graphicx}
\usepackage[update,prepend]{epstopdf}
\usepackage{hyperref}
\usepackage{array}
\usepackage{cite}
\usepackage{natbib}
\usepackage{morefloats}
\usepackage{longtable,tabularx}

\begin{document}

\title{Analysis of Scattering from Archival Pulsar Data using a CLEAN-based Method}

\author{
Jr-Wei Tsai\altaffilmark{1},
John H. Simonetti\altaffilmark{1},
Michael Kavic\altaffilmark{2}
}

\altaffiltext{1}{Department of Physics, Virginia Tech, Blacksburg, VA 24061, U.S.A}
\altaffiltext{2}{Department of Physics, Long Island University, Brooklyn, New York 11201, U.S.A.}

\begin{abstract}
In this work we adopted a CLEAN-based method to determine the scatter time, $\tau$, from archived pulsar profiles under both the thin screen and uniform medium scattering models and to calculate the scatter time frequency scale index $\alpha$, where $\tau\propto\nu^{\alpha}$.  The value of $\alpha$ is $-4.4$, if a Kolmogorov spectrum of the interstellar medium turbulence is assumed. We deconvolved 1342 profiles from 347 pulsars over a broad range of frequencies and dispersion measures. In our survey, in the majority of cases the scattering effect was not significant compared to pulse profile widths. For a subset of 21 pulsars scattering at the lowest frequencies was large enough to be measured. 
Because reliable scatter time measurements were determined only for the lowest frequency, we were limited to using upper limits on scatter times at higher frequencies for the purpose of our scatter time frequency slope estimation.
We scaled the deconvolved scatter time to 1~GHz assuming $\alpha=-4.4$ and considered our results in the context of other observations which yielded a broad relation between scatter time and dispersion measure.
\end{abstract}


\section{Introduction}
Scattering due to the interstellar medium (ISM) is often assumed to be described by a Kolmogorov spectrum which is consistent with pulsar observations of angular broadening and scintillation (e.g. \citealt{1995ApJ...443..209A} for distances $\lesssim$ 1~kpc).  However, there are also observations of pulsars which exhibit a departure from the Kolmogorov spectrum.  Studies of pulsars with large dispersion measures (DMs) by \citet{2004ApJ...605..759B} show a flattened scattering spectral index. This analysis was performed by combining the CLEAN-based algorithm at high frequencies and the converted scatter time from correlative bandwidths caused by scintillation at low frequencies. It has been suggested that the effect of scattering for some pulsars is highly dependent on the line-of-sight (LOS) \citep{1985ApJ...288..221C,2004A&A...425..569L,2015MNRAS.449.1570L}

These results suggest it may prove futile to attempt to formulate a general relation of scattering to frequency and DM (e.g. \citet{2015MNRAS.449.1570L}). A comprehensive study of the effect of scattering on the emission from a large number of pulsars at various DMs and with different LOSs could be used to clarify this issue. A critical amount of archival pulsar data is available for such an analysis. Here we report results from a survey of the scattering effect of 1342 pulsar profiles from 347 pulsars that came from the LWA Pulsar Data Archive \citep{2014arXiv1410.7422S} and European Pulsar Network (EPN) data archive.

The majority of profiles do not show a significant scattering tail.  Thus the deconvolved exponential decay time of the majority of the pulsars provides an upper limit on the scatter time. A subset of 21 pulsars showed a significant scattering tail in the lowest frequencies compared with higher frequencies. Because we did not find any pulsars for which two recorded profiles at different frequencies both showed a significant scattering tail, it remains difficult to interpret an off-Kolmogorov spectrum of the free electron turbulence along the LOS from our results.

Assuming an scatter time frequency scale index $\alpha=-4.4$ from
\begin{equation}
\tau \propto \nu^{\alpha},
\end{equation}
where $\tau$ is the scatter time and $\nu$ is the frequency, we scaled the deconvolved scatter time assuming the thin screen scattering model to the frequency of 1~GHz and placed our results in the context of other studies. Our results support a broad relation between the scatter time and DM.  The scatter time value is as wide as two orders of magnitude for a given range of DM, and the range could be wider for DMs larger than 100 pc cm$^{-3}$. %Although we did not find any pulsar which shows significant scattering tails at two different frequencies, we were still able to calculate the upper limits for the $\alpha$ of the 21 pulsars.
%, because we may overestimate the scatter time at higher frequencies. 
We found a mean value of upper limit of $\lesssim -3.8$ for the thin screen scattering model from a subset of 21 pulsars.

\section{Background}\label{sect:scatter}
Spatial variations in the interstellar free-electron number density are responsible for the scattering and scintillation of radio signals propagating through the ISM.  Pulsar observations are particularly useful for measuring the effects of interstellar scattering, and thus can be used to characterize the interstellar electron-density irregularities. 

Analysis of interstellar scattering often uses a thin screen between the source and observer. \citet{1986MNRAS.220...19R} showed the range of the viable spatial electron density spectra is virtually limited to the (approximately) power-law spectra with the index ranging from 11/3 to 4.3
%This implies electron number-density fluctuations with a power-law spatial power spectrum \citep{1977ARA&A..15..479R}.  
For a wavenumber $q$ between the outer and inner scales of the irregularities, $q_o$ and $q_i$, i.e., $q_o \ll q \ll q_i$, the power-spectrum for the electron-density irregularities can be expressed as a power-law,
\begin{equation}
P_{n_e}=C^2_{n}{q^{-\beta}},
\end{equation}
where $C_{n}$ is the fluctuation strength for a given LOS. Generally, for a power-law wavenumber spectrum, the broadening time $\tau$ follows a power law,
\begin{equation}
\tau \propto \nu^{\alpha} \propto \nu^{2\beta/(\beta-2)}.
\label{scale}
\end{equation}
For a Kolmogorov spectrum $\beta = 11/3$, the $\alpha$ is $-4.4$ (e.g., see \citet{2004ApJ...605..759B}).

There are observations of pulsars over a wide range of DMs which exhibit a departure from the thin screen model with Kolmogorov's density fluctuation spectrum. For example, a small group of pulsars with a high DM range (582-1074 pc cm$^{-3}$) were observed to have an average $\alpha = -3.44 \pm 0.13$ \citep{2001ApJ...562L.157L}) and $\alpha = 3.49$ for DM $>$ 500 pc cm$^{-3}$ \citep{2015MNRAS.449.1570L}. A larger group of low galactic-latitude pulsars were observed to have an average $\alpha = -3.9 \pm 0.2$ \citep{2004ApJ...605..759B}.

Since several studies of the scattering spectral index \citep{2001ApJ...562L.157L, 2004ASSL..315..327L, 2004ApJ...605..759B, 2013MNRAS.434...69L, 2015MNRAS.449.1570L} revealed that only a handful of pulsars have their scaling indices close to the theoretical value of $\alpha$. \citet{2004ApJ...605..759B} and \citet{2013MNRAS.434...69L} explored several plausible explanations for the departure from $\alpha =-4.4$. In order for $\alpha=-4.4$ all four of the following conditions must be fulfilled. (i) The electron density spectrum is of the Kolmogorov form. (ii) Only a thin screen or a uniformly thick bulk lies in the LOS. (iii) The wavenumbers sampled by the observations fall in the range between the inner and outer scales. (iv) The turbulence is isotropic and homogeneous. Thus there are a variety of deviations from these conditions which could explain the departure from $\alpha=-4.4$.  Some pulsar observations which deviate from Kolmogorov results have been attributed to deviations from the assumptions summarized above. For example some atypical geometries, such as the truncated screens proposed by \citet{2001ApJ...549..997C} may also affect the observed slope of the scatter time-frequency relation. It has also been proposed that observations on a spatial scale smaller than the inner turbulence scale can cause a deviation from Kolmogorov resulting in an $\alpha>-4.4$ \citep{1986MNRAS.220...19R}.

\section{Observations and Data Reduction}\label{sect:data}
As discussed above, pulsars provide a great tool to probe the effect of scattering along different LOSs, with different frequencies and DMs.  We conducted a survey of the effect of scattering which included 1342 profiles from 347 pulsars to determine the best fit scatter time in the thin screen and uniform medium scattering models. The pulsar profiles used were drawn from from the LWA Pulsar Data Archive \citep{2014arXiv1410.7422S} and European Pulsar Network (EPN) data archive\footnote{\tt http://www.jb.man.ac.uk/pulsar/Resources/epn/}. The analysis covers a frequency range from 25~MHz to 43~GHz and DMs from 2.38 to 780 pc cm$^{-3}$.

We made use of the CLEAN-based algorithm \citep{2003ApJ...584..782B} to deconvolve the scattering effect on the pulsar profiles.  Each recorded pulsar profile is composed of the averaged intrinsic pulse convolved with propagation effects and instrumental responses.  The CLEAN-based algorithm utilizes an accumulated delta-like signal to restore the intrinsic pulse.  This approach allows for the deconvolution of various profile shapes without knowledge of the intrinsic profile.
%It should be noted that the CLEAN-based method does not require knowledge about the intrinsic pulse profile. 
It does, however, makes an important assumption; it assumes that all scatter-like features in the profile (i.e. an asymmetry that makes the right hand side slope of the profile flatter than the left) are indeed coming from scattering.

So the recorded signal $P_{obs}$ is
\begin{equation} 
P_{obs}(t) = I(t)\otimes {\rm PBF}(t)\otimes r(t),
\label{convolutions}
\end{equation}
where $I(t)$ is the intrinsic pulse profile,  PBF$(t)$ is the pulse-broadening function, and $r(t)$ is a response function which gives the combined instrumental responses including effects due to data reduction. 

We deconvolved the profile with two scattering models, the thin screen and the uniform medium models. As their names suggest these models make different assumptions about the distribution of the scattering medium between the source and the observer \citep{1972MNRAS.157...55W,1973MNRAS.163..345W}.
For the majority of the distant pulsars the thin screen model provides a good approximation of effect of scattering. The amount of scattering depends on the magnitude of density fluctuations in the ionized ISM. The fluctuations are obviously greatest in high electron density regions, such as supernova remnants or HII regions. Within these regions the electron density may be a few orders of magnitude larger than in an average ISM region. Thus the density fluctuations will be larger as well. When one considers that the size of such regions is at most a few tens-of-parsecs, compared to the kilo-parsec distances to many pulsars, the concept of a thin screen seems appropriate. The situation is a bit different for nearby pulsars, where the existence of such scattering screens along the line of sight may be unlikely. Thus we also analyze the effect of scattering using the ``uniform medium'' model. 
%which is opposite to thin screen model, 
As the name suggested this model assumes the density fluctuations are spread uniformly along the entire LOS.

The normalized pulse-broadening functions for these two models are given by
\begin{eqnarray}
{\rm PBF}_{\rm ts}(t) &=& \tau_{\rm ts}^{-1}{\rm exp}(-t/ \tau_{\rm ts}) U(t)
\label{thinscreen}\\
{\rm PBF}_{\rm um}(t) &=& (\pi^5\tau_{\rm um}^3/16t^5)^{1/2}{\rm exp}(-\pi^2\tau_{\rm um}/4t)U(t)
\label{unimedium}
\end{eqnarray}
where PBF$_{\rm ts}$ and $\tau_{\rm ts}$ are the pulse-broadening function and scatter time for the thin screen model, and PBF$_{\rm um}$ and $\tau_{\rm um}$ are the pulse-broadening function and scatter time for the uniform medium model, $U(t)$ is a unit step function where $U(< 0) = 0$ and $U(\ge 0) = 1$.

The choice of $r(t)$ effects the shape of the restored profile and may effect the scatter time measurement. $r(t)$ usually is chosen to be a Gaussian because $r(t)$ is a convolution of several step functions \citep{2003ApJ...584..782B}. To choose a proper FWHM of $r(t)$, we first consider that some of the older profiles from the EPN database had poor time resolution yielding time constants that were larger than the sampling time and resulting in a broader instrumental response function. Also, dispersion smearing was a factor in data reduction as only incoherent dedispersion was applied and at low frequencies of the order of a few tens of MHz.  This was the case for most of the older profiles included in the EPN database, since most of these observations were made using wide-band filterbank receivers.  Thus in some cases where $r(t)$ is assumed to be a Gaussian with a small FWHM the restored profiles can contain a collections of narrow spikes. Here we simply assumed $r(t)$ to be a Gaussian with a FWHM that was equal to the $4\times$temporal bin width and iteratively subtracted the trial PBF$(t)\otimes r(t)$ from $P_{obs}(t)$.

We tried scatter times which ranged from $0.2$ temporal bin widths to the observed profile width. In each trial we iterated by subtracting scattered bins as discussed in \citet{2003ApJ...584..782B} until the resulting average of the profile was smaller than the average of the time series outside the profile, or the standard deviation of the profile was smaller than the standard deviation of the time series outside the profile. We thus identified the scatter time as the time constant which had the smallest standard deviation ratio between the outside and inside of the profile.

The CLEAN-based method can be used to extract information about the effect of scattering only if one assumes that the intrinsic pulsar profile does not contain other features that may resemble scattering. This is also the case for other scatter time estimation methods \citep{2013MNRAS.434...69L}. If we imagine that the pulsar has an intrinsically asymmetric profile which is not scattered but which contains a feature that mimics the appearance of a scattering tail (see the profiles of the Vela pulsar at $>$ 1~GHz for example), when the CLEAN method (or any other) is applied to extract the scatter time from such a profile a non-zero value of $\tau$ will be found. %This is because these methods are purely mathematical.
As \citet{2013MNRAS.434...69L} and \citet{2015ApJ...804...23K} argued one can obtain reliable values of $\tau$ only when the scatter time is significantly larger than the width of the profile at higher frequency.

When we apply this method to a given profile the value of $\tau$ derived is most likely a result of evolution of the profiles at different frequencies dominated by effects unrelated to scattering. In such cases $\tau$ shows some kind of frequency evolution of this ``fake" scatter time.  This fake-scatter tail will lengthen, yielding larger values of the fake-pulse broadening time at lower frequencies.
%When the spectral index is calculated those broadening effect, not scattering, will result in a larger value for $\alpha$, where $\tau \propto \nu^{\alpha}$.
Since most of the profiles become broader at lower frequencies (due to radius-to-frequency mapping), this fake scatter time will in most cases also increase, mimicking the ``real'' scattering evolution, although usually with much flatter slope, yielding in turn fake values of $\alpha \gg -4.4$.

\section{Results} \label{sect:Results}
We carefully examined all profiles from each pulsar at all available observing frequencies from the EPN and LWA Pulsar Data Archive. We found a subset of 21 pulsars which significantly showed a scattering tail in the profile at lower frequencies compared with higher frequencies. From the subset of 21 pulsars we assume that the scattering effect dominated the profile tail and deconvolved the effect making use of the two scatter scenarios: the thin screen and uniform medium models, as shown in figures~\ref{ts} and \ref{am}. The individual scatter times are shown in the table~\ref{tau}. 
For the sake of comparison we list the values of $\tau_{\rm ts}$ at similar frequencies from previous works~\citep{2001ApJ...562L.157L, 2013MNRAS.434...69L, 2015MNRAS.454.2517L} in table~\ref{tab:compare}.
%To place the result described here in context we 
We scaled $\tau_{\rm ts}$ to 1~GHz assuming a $\alpha=-4.4$ along with other works \citep{2003astro.ph..1598C,2004ApJ...605..759B} in figure~\ref{scaledtau}. Note that our results include three new direct scatter time measures of low DM ($<100$ pc cm$^{-3}$) pulsars, which are not converted from measurements of the decorrelation bandwidth. Other values of $\tau_{\rm ts}$ shown in figure~\ref{scaledtau} which were not calculated from the lowest available frequency of the subset of 21 pulsars 
%serve as upper limits that combined scattering with other effects as discussed above.
were used to estimate scatter time upper limits. Our results support a broad relation between scatter time and DM. In figure \ref{scaledtau}, we can see that the scatter time generally spans two orders of magnitudes for DM $\lesssim 100$ pc cm$^{-3}$, and spans over even larger orders for DM $\gtrsim$ 100, within a range of DM.

The scatter time index, $\alpha$ can be calculated for observations at two different frequencies by making use of the following relation
%ing %equation~\ref{scalelogform}
\begin{equation}
\alpha = \frac{{\rm log}\ \tau_{{\rm high}}-{\rm log}\ \tau_{\rm low} }{ {\rm log}\ \nu_{\rm high} - {\rm log}\ \nu_{\rm low}  },
\label{scalelogform}
\end{equation}
where we make use of the two respective scatter times and observing frequencies. Given that the scattering effect is not observed in the next to the lowest available frequency of the 21 pulsars, we can make use of the above expression to set an upper limit on $\alpha$ which we list them in table~\ref{tau}. The mean values of the upper limit on $\alpha$ are $-3.8$ and $-4.8$ (from 13 pulsars) from the thin screen and the uniform medium scattering models, respectively.

We excluded results using the uniform medium scattering model for a subset of 21 the pulsars because this model is not appropriate for those profiles. This conclusion is based on the fact that the deconvolved results still show an obvious scattering tail or a collection of narrow spikes that are not expected in a natural pulsar profile.

Note that we deconvolved the scatter time from the profile assuming an intrinsic profile composed of Gaussian components. However, this will not necessarily be the case for all intrinsic pulsar profiles. Thus the departures from the Kolmogorov scaling index we found for some pulsars should not necessarily be inferred as anomalies of the free electron turbulence along the LOSs.

\section{Conclusion}\label{sect:conclusion}

We used the CLEAN-based method to deconvolve the effect of scattering for 1342 profiles from 347 pulsars to derive the spectral index and time constant within a broad range of frequencies and DM, assuming the thin screen and uniform medium scattering models. Many scatter times were not significant compared to the profile widths but nonetheless provide bounds on the effect of scattering at higher frequencies. 

A subset of 21 pulsars showed significant effects of scattering at the lowest available frequency, and were used to calculate upper limits on $\alpha$. We found a mean value of upper limit of $\alpha \lesssim -3.8$ using the thin screen scattering model from 21 pulsars. We also calculated a mean value of upper limit of $\alpha$ as $\lesssim -4.8$ using the uniform medium scattering model from 13 pulsars. Our results include three new direct scatter time measurements of low DM ($<$100 pc cm$^{-3}$) pulsars.

It is difficult to infer an off-Kolmogorov spectrum of the free electron turbulence along the LOS from our results. The scaled deconvolved time constant supports broad relations between the DM and scatter time, which can spans over two orders of magnitudes given a range of DMs.

\begin{figure*}
\begin{center}
\includegraphics[width=1\textwidth]{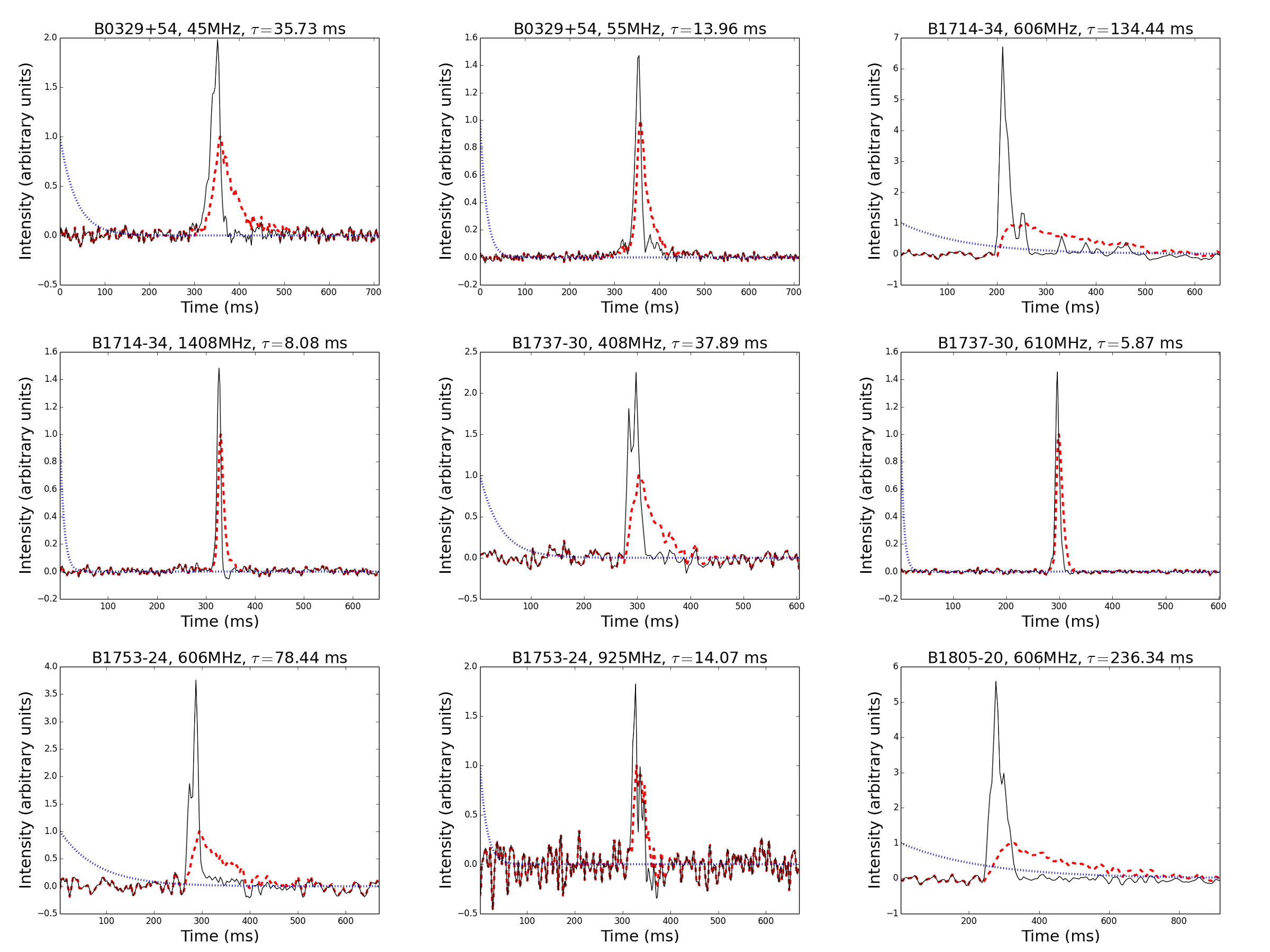}
\caption{
The deconvolved scatter time and clean components of 21 pulsars assuming the thin screen model is shown. The red thick dashed line is the average profile of a given pulsar. The blue thin dashed line is the curve of the scattering function. The black solid line is the clean components with noise deconvolved from the profile. Clean components mathematically represent the intrinsic profile before being scattered but due to poor temporal resolution after data reduction certain unphysical features such as
%SUCH AS WHAT??? 
collections of narrow spikes appear in some clean components.
}
\label{ts}
\end{center}
\end{figure*}

\addtocounter{figure}{-1}
\begin{figure*}
\begin{center}
\includegraphics[width=1.\textwidth]{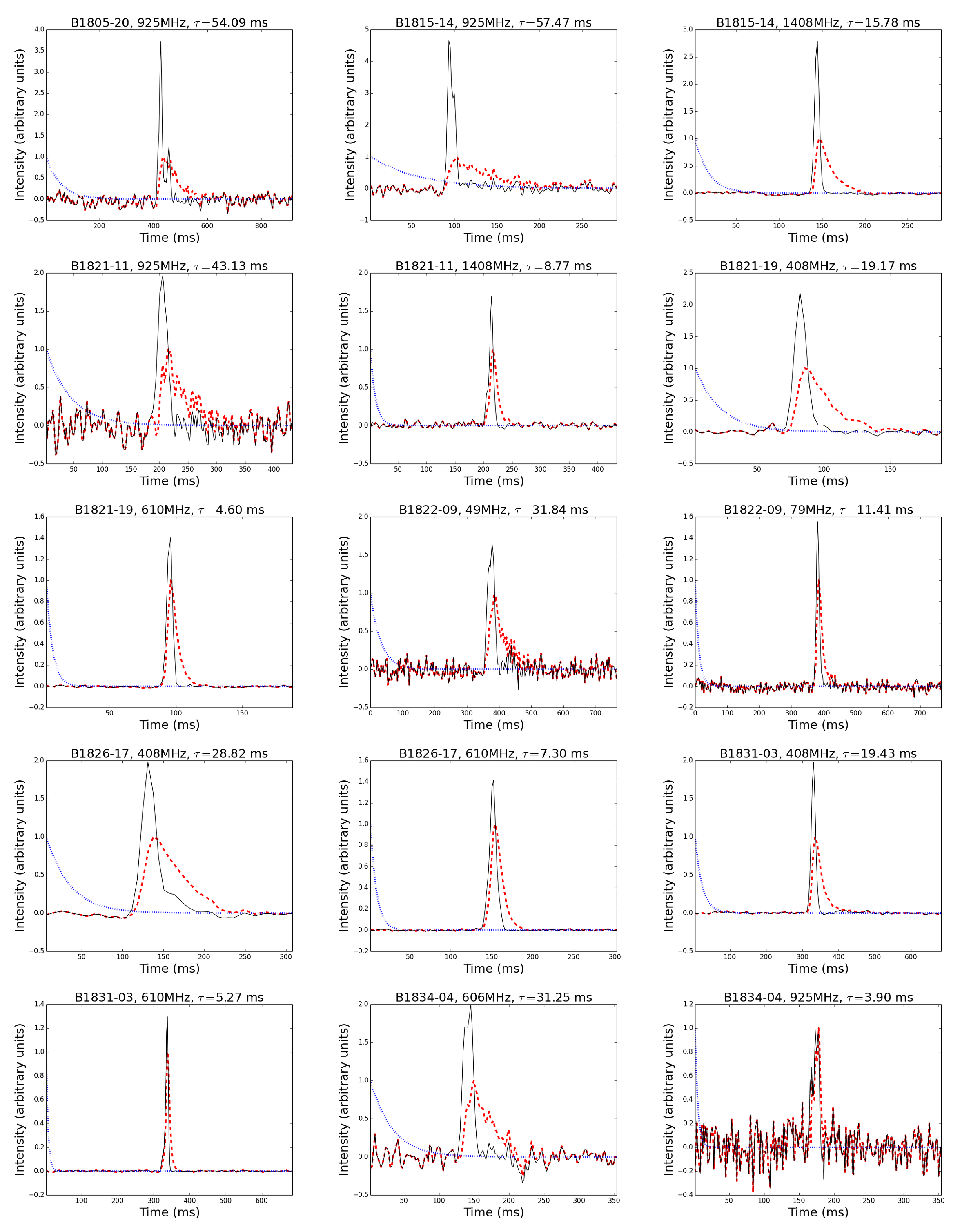}
\caption{Cont.}
\end{center}
\end{figure*}

\addtocounter{figure}{-1}
\begin{figure*}
\begin{center}
\includegraphics[width=1.\textwidth]{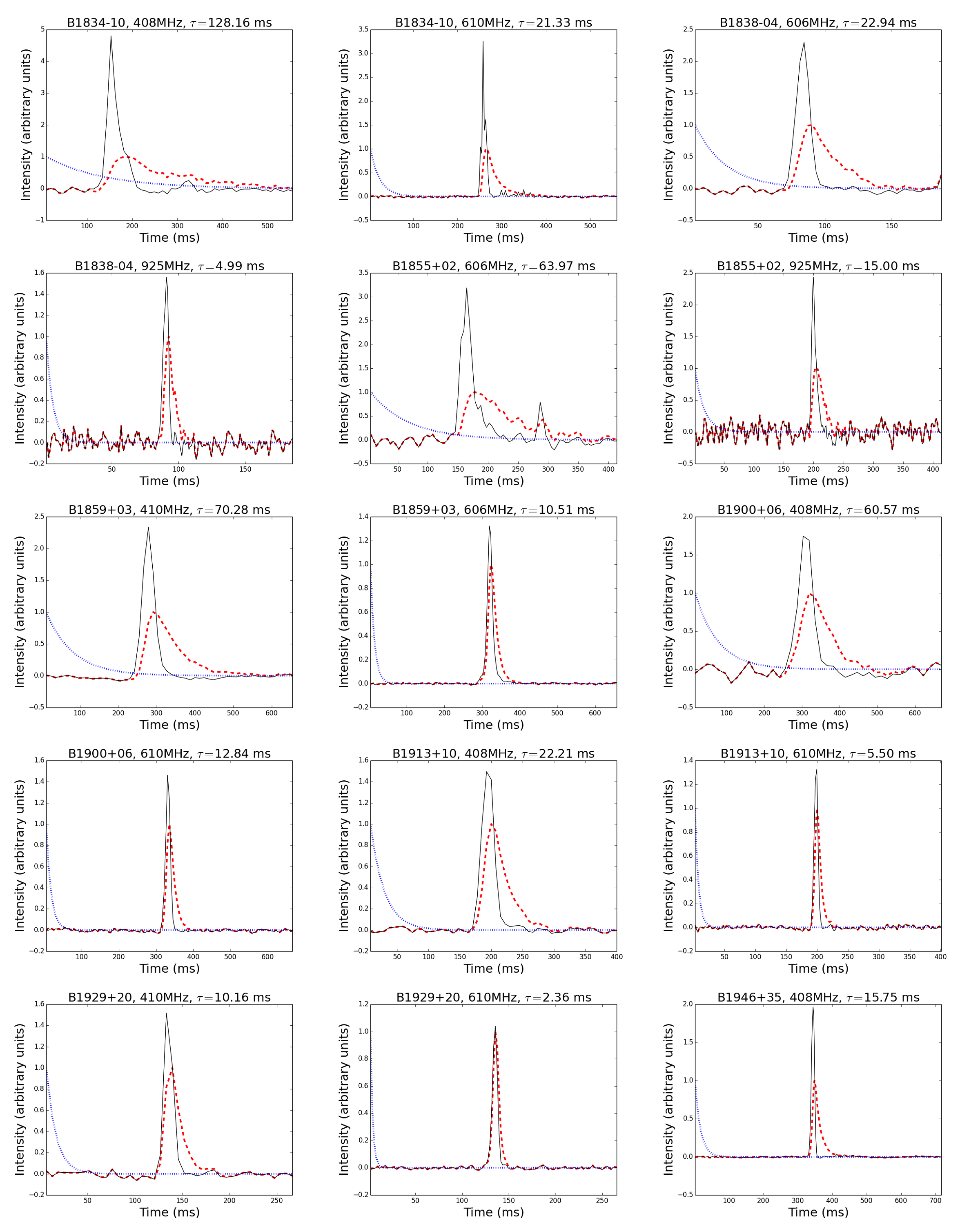}
\caption{Cont.}
\end{center}
\end{figure*}

\addtocounter{figure}{-1}
\begin{figure*}[ht!]
\begin{center}
\includegraphics[width=1.\textwidth]{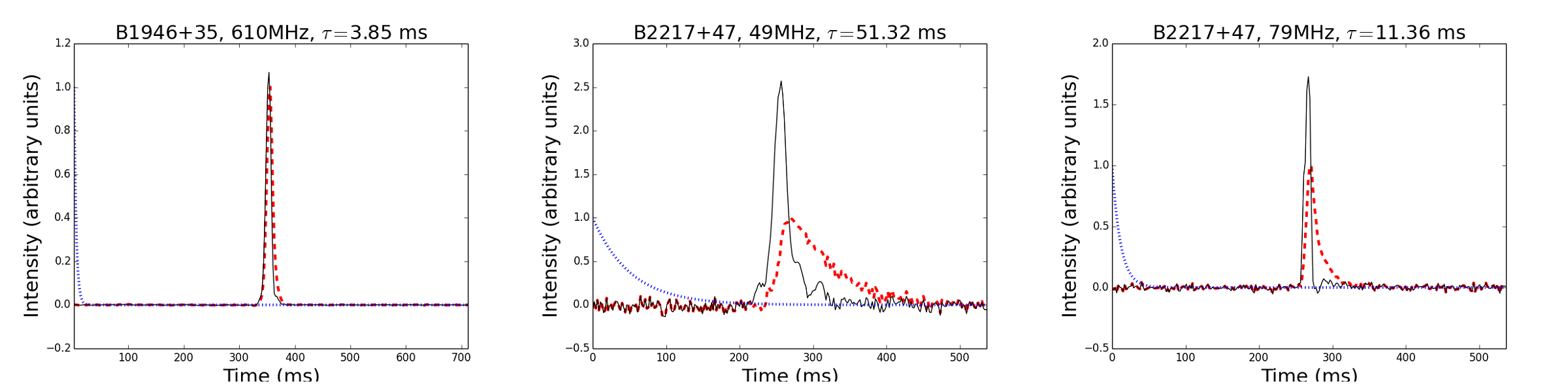}
\caption{Cont.}
\end{center}
\end{figure*}

\begin{figure*}
\begin{center}
\includegraphics[width=1\textwidth]{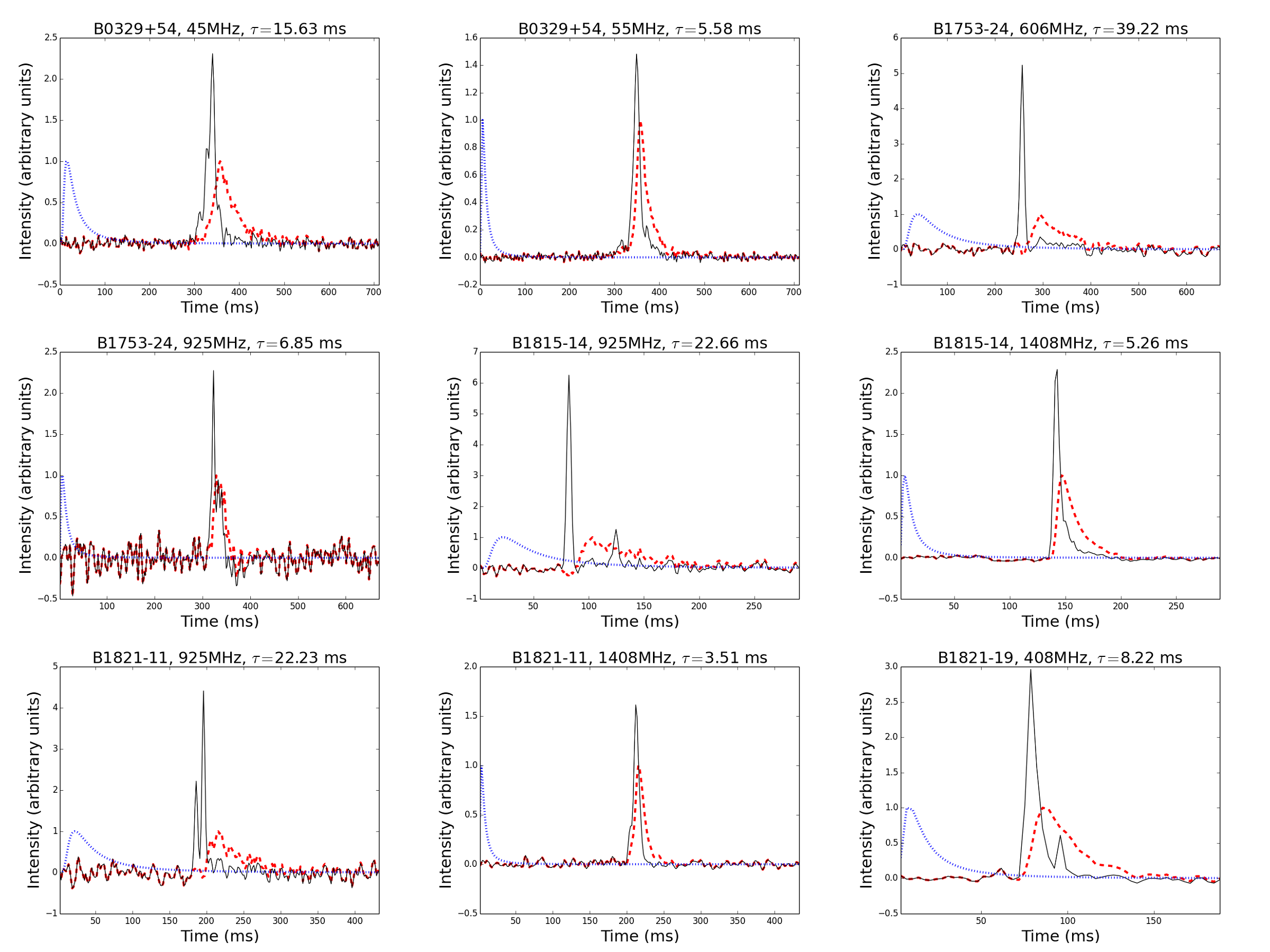}
\caption{
The deconvolved scatter time and clean components of 13 pulsars assuming the uniform medium scattering model is shown. The red thick dashed line is the average profile of a given pulsar. The blue thin dashed line is the curve of the scattering function. The black solid line is the clean components with noise deconvolved from the profile. Clean components mathematically represent the intrinsic profile before being scattered but due to poor temporal resolution after data reduction certain unphysical features such as collections of narrow spikes appear in some clean components.
}
\label{am}
\end{center}
\end{figure*}

\addtocounter{figure}{-1}
\begin{figure*}
\begin{center}
\includegraphics[width=1.\textwidth]{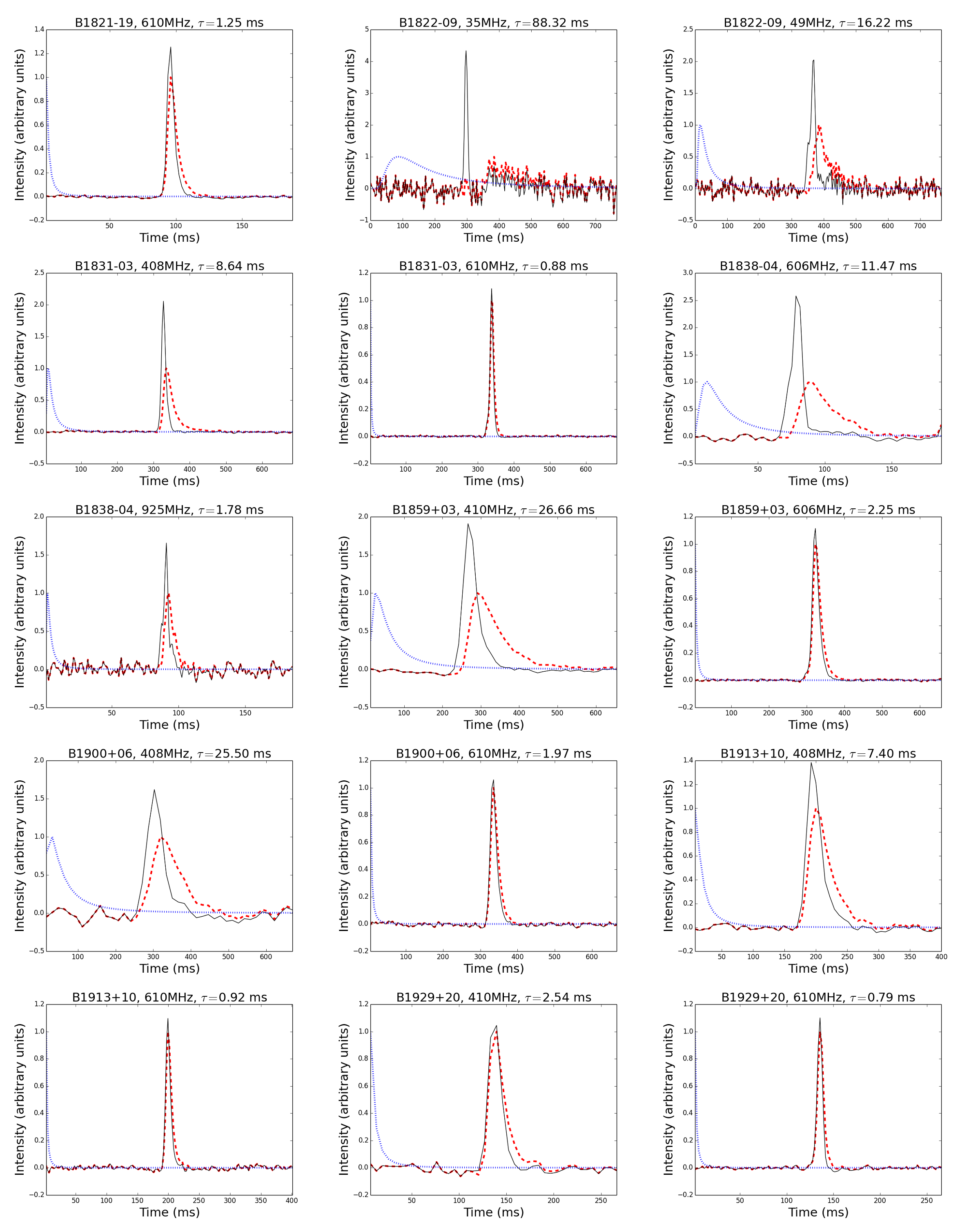}
\caption{Cont.}
\end{center}
\end{figure*}

\addtocounter{figure}{-1}
\begin{figure*}
\begin{center}
\includegraphics[width=1.\textwidth]{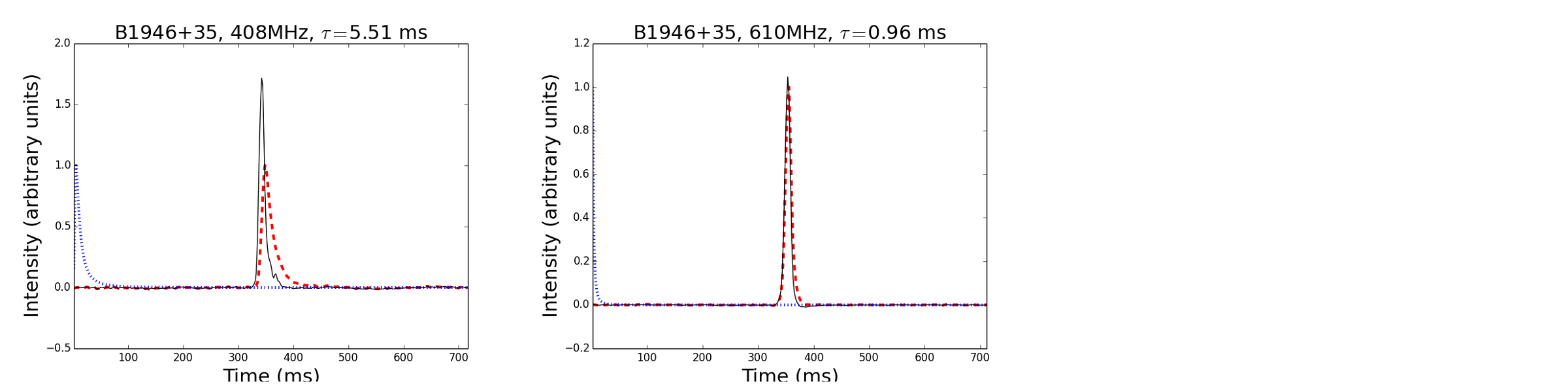}
\caption{Cont.}
\end{center}
\end{figure*}

\begin{figure*}[ht!]
\begin{center}
\includegraphics[width=.75\textwidth]{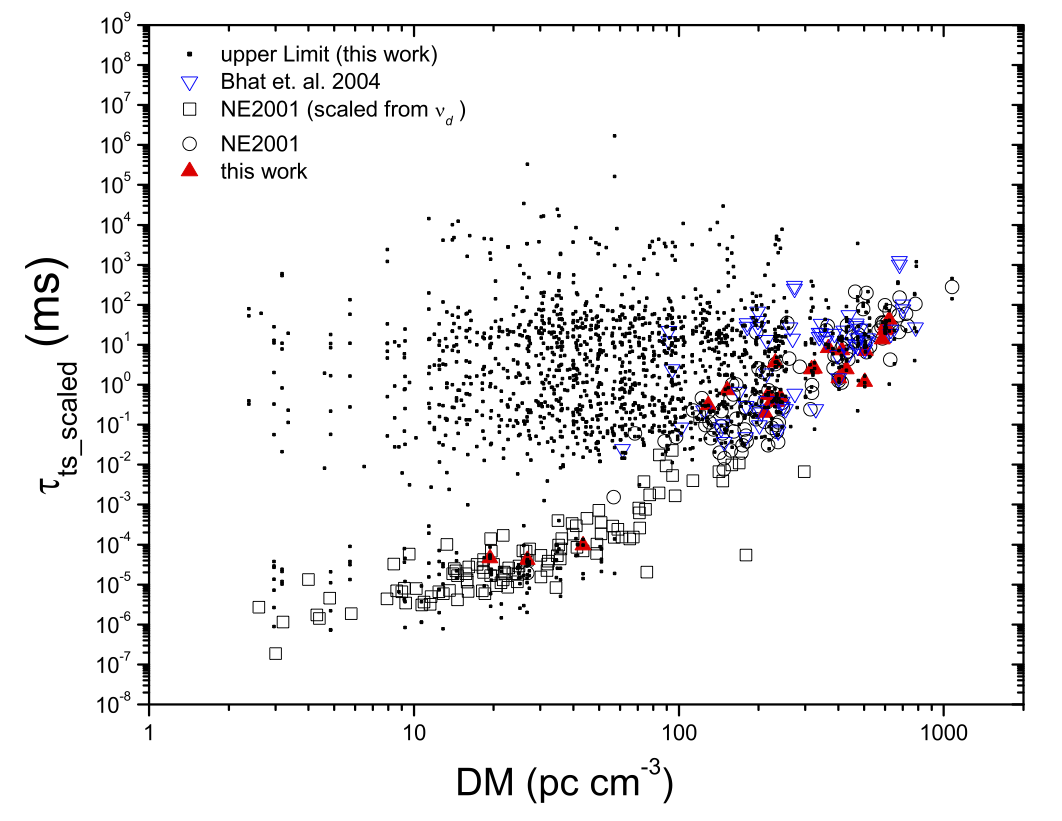}
\caption{The scaled $\tau_{\rm ts}$ at 1~GHz assuming $\alpha=-4.4$, where $\tau\propto\nu^\alpha$.  Some of $\tau_{\rm ts}$ are converted from $\nu_d$ by $\tau\nu_d =  \rm C_1/2\pi$ , assuming $\rm C_1=1$.
}
\label{scaledtau}
\end{center}
\end{figure*}

\begin{center}
\begin{longtable}{lccccccc}
\caption{The scatter time and the upper limit value of $\alpha$ ($\alpha$) from the CLEAN-base method. The subscript ``ts" and ``um" indicate the thin screen scattering model and uniform medium scattering model, respectively. Note that for some of the pulsars the $\tau_{\rm um}$ is not available because the phases of deconvolved peak components are outside of the range of the profile due to limitations of our data reduction algorithm. The error is the step size during the deconvolution trial. ($^*$Archive data from LWA Pulsar Data Archive, others are from EPN. $^{**}$The CLEAN-base method is not valid due to improper clean bin phase position.)
} 
\label{tau}\\
\hline \multicolumn{1}{c}{\textbf{PSR}} &
\multicolumn{1}{c}{\textbf{$\nu$ (MHz)}} &
\multicolumn{1}{c}{\textbf{$\tau_{\rm ts}$ (ms)}} &
\multicolumn{1}{c}{\textbf{$\tau_{\rm um}$ (ms)}} &
\multicolumn{1}{c}{\textbf{error (ms)}} &
\multicolumn{1}{c}{\textbf{$\alpha_{\rm ts}$}} &
\multicolumn{1}{c}{\textbf{$\alpha_{\rm um}$}} \\ \hline 
\endfirsthead

\multicolumn{7}{c}%
{{\bfseries \tablename\ \thetable{} -- continued from previous page}} \\
\hline \multicolumn{1}{c}{\textbf{PSR}} &
\multicolumn{1}{c}{\textbf{$\nu$ (MHz)}} &
\multicolumn{1}{c}{\textbf{$\tau_{\rm ts}$ (ms)}} &
\multicolumn{1}{c}{\textbf{$\tau_{\rm um}$ (ms)}} &
\multicolumn{1}{c}{\textbf{error (ms)}} &
\multicolumn{1}{c}{\textbf{$\alpha_{\rm ts}$ }} &
\multicolumn{1}{c}{\textbf{$\alpha_{\rm um}$}} \\ \hline 
\endhead

\hline \multicolumn{7}{r}{{Continued on next page}} \\ \hline
\endfoot

\hline \hline
\endlastfoot

B0329+54$^*$ & 45 & 35.73 & 15.63 & 0.56 & $<-$4.68 & $<-5.13$ \\
B0329+54$^*$ & 55 & $<13.96$ & $<5.58$ & 0.56 & & \\  
B1714$-$34 & 606 & 134.44 & & 1.08 & $<-$3.34 & \\  
B1714$-$34 & 1408 & $<8.08$ & & 0.35 & & \\   
B1737$-$30 & 408 & 37.89 & & 0.9 & $<-$4.64 & \\  
B1737$-$30 & 610 & $<5.87$ & & 0.28 & & \\   
B1753$-$24 & 606 & 78.44 & 39.22 & 0.68 & $<-$4.06 & $<-4.13$ \\
B1753$-$24 & 925 & $<14.07$ & $<6.85$ & 0.38 & & \\  
B1805$-$20 & 606 & 236.34 & & 1.13 & $<-$3.67 & \\  
B1805$-$20 & 925 & $<54.09$ & & 0.67 & & \\   
B1815$-$14 & 925 & 54.47 & 22.66 & 0.33 & $<-$2.95 & $<-3.48$ \\
B1815$-$14 & 1408 & $<15.78$ & $<5.26$ & 0.38 & & \\  
B1821$-$11 & 925 & 43.13 & 22.23 & 0.33 & $<-$3.79 & $<-4.39$ \\
B1821$-$11 & 1408 & $<8.77$ & $<3.51$ & 0.35 & & \\  
B1821$-$19 & 408 & 19.17 & 8.22 & 0.68 & $<-$3.85 & $<-4.68$ \\
B1821$-$19 & 610 & $<4.06$ & $<1.25$ & 0.42 & & \\  
B1822$-$09$^*$ & 35 & 116.55 & 88.32 & 0.6 & $<-$3.86 & $<-5.04$ \\
B1822$-$09$^*$ & 49 & $<31.84$ & $<16.22$ & 0.6 & & \\  
B1826$-$17 & 408 & 28.82 & & 1.31 & $<-$3.41 & \\  
B1826$-$17 & 610 & $<7.3$ & & 0.41 & & \\   
B1831$-$03 & 408 & 19.43 & 8.64 & 0.72 & $<-$3.24 & $<-5.68$ \\
B1831$-$03 & 610 & $<5.27$ & $<0.88$ & 0.44 & & \\  
B1834$-$04 & 606 & 31.25 & & 0.43 & $<-$4.92 & \\  
B1834$-$04 & 925 & $<3.9$ & & 0.24 & & \\   
B1834$-$10 & 408 & 128.16 & & 1.91 & $<-$4.46 & \\  
B1834$-$10 & 610 & $<21.33$ & & 0.59 & & \\   
B1838$-$04 & 606 & 22.94 & 11.47 & 0.6 & $<-$3.61 & $<-4.41$ \\
B1838$-$04 & 925 & $<4.99$ & $<1.78$ & 0.18 & & \\  
B1855+02 & 606 & 63.97 & & 0.94 & $<-$3.43 & \\  
B1855+02 & 925 & $<15$ & & 0.28 & & \\   
B1859+03 & 410 & 70.28 & 26.66 & 2.42 & $<-$4.86 & $<-6.33$ \\
B1859+03 & 606 & $<10.51$ & $<2.25$ & 0.75 & & \\  
B1900+06 & 408 & 60.57 & 25.5 & 3.19 & $<-$3.86 & $<-6.37$ \\
B1900+06 & 610 & $<12.84$ & $<1.97$ & 0.99 & & \\  
B1913+10 & 408 & 22.21 & 7.4 & 1.48 & $<-$3.47 & $<-5.18$ \\
B1913+10 & 610 & $<5.5$ & $<0.92$ & 0.46 & & \\  
B1929+20 & 410 & 10.16 & 2.54 & 1.27 & $<-$3.67 & $<-2.94$ \\
B1929+20 & 610 & $<2.36$ & $<0.79$ & 0.39 & & \\  
B1946+35 & 408 & 15.75 & 5.51 & 0.39 & $<-$3.5 & $<-4.34$ \\
B1946+35 & 610 & $<3.85$ & $<0.96$ & 0.48 & & \\  
B2217+47$^*$ & 49 & 51.32 & & 0.42 & $<-$3.16 & \\  
B2217+47$^*$ & 79 & $<11.36$ & & 0.42 & & \\
\end{longtable}
\end{center}

\begin{table}
\begin{center}
\caption{This table shows the comparison of scatter time from this work and previous works. ($^1$\citet{2013MNRAS.434...69L}, $^2$\citet{2015MNRAS.454.2517L}, $^3$\citet{2001ApJ...562L.157L})
}
\label{tab:compare}
%\begin{tabular}{c|c|c|c|c}
\begin{tabular}{lccccccccc}
\hline
\multicolumn{1}{c}{PSR} & & & 
\multicolumn{2}{c}{this work} & & &
\multicolumn{2}{c}{previous works}\\
\cline{4-5} \cline{8-9}
%\hline
B1815$-$14 & & & 1.4 GHz & 15.78$\pm$0.38  & & & 1.4 GHz & 15.3$\pm$2.0$^1$ \\
B1834$-$04 & & & 606 MHz & 31.25$\pm$0.43  & & & 610 MHz & 33$\pm$5$^1$ \\
B1929+20   & & & 610 MHz & 2.36$\pm$0.39   & & & 610 MHz & 1.71$\pm$0.21$^2$ \\
B1805$-$20 & & & 606 MHz & 236.34$\pm$1.13 & & & 0.6 GHz & 259$\pm$47$^3$ \\
B1821$-$11 & & & 925 MHz & 43.13$\pm$1.34  & & & 0.9 GHz & 40$\pm$12$^3$ \\
\end{tabular}
\end{center}
%\label{tab:compare}
\end{table}

\end{document}